\documentclass[11pt]{article}

\usepackage{amsfonts,latexsym,eucal,color,graphicx,epsfig,amssymb,amsmath}

\newcommand{\be}{\begin{eqnarray}}
\newcommand{\ee}{\end{eqnarray}}
\newcommand{\pd}{\partial}

\newcommand{\dd}{\mathrm{d}}
\newcommand{\dalm}{\kern1pt\vbox{\hrule height 0.9pt\hbox{\vrule width 0.9pt\hskip 2.5pt\vbox{\vskip 5.5pt}\hskip 3pt\vrule width 0.3pt}\hrule height 0.3pt}\kern1pt}

\begin{document}


\begin{flushright}
{\small
{\tt arXiv:0908.1019 [hep-th]}\\
 WU-AP/303/09\\
}
\end{flushright}

\vspace{.4cm}

\begin{center}
{\Large {\bf Stability of branes trapped by $d$-dimensional black holes}}
\end{center}

\vspace{.2cm}

\begin{center}
{\large
Kenta Hioki${}^{a, b,}$\footnote{{\tt hioki@gravity.phys.waseda.ac.jp}},
Umpei Miyamoto${}^{c,}$\footnote{{\tt umpei@phys.huji.ac.il}},
and Masato Nozawa${}^{a,}$\footnote{{\tt nozawa@gravity.phys.waseda.ac.jp}}
}

\vspace{.6cm}

${}^{a}${\it Department of Physics, Waseda University,\\Okubo 3-4-1, Tokyo 169-8555, Japan}\\
\vspace{3mm}
${}^{b}${\it Waseda Research Institute for Science and Engineering,\\Okubo 3-4-1, Tokyo 169-8555, Japan}\\
\vspace{3mm}
${}^{c}${\it Racah Institute of Physics, Hebrew University,\\Givat Ram, Jerusalem 91904, Israel}\\

\end{center}

\vspace{.3cm}

\begin{abstract}
The system of extended objects interplaying with a black hole describes
 or mimics various  gravitational phenomena. In this brief paper, we
 report the results of stability analysis of codimension-one
 Dirac-Nambu-Goto branes rest at the equatorial plane of $d$-dimensional
 spherical black holes, including the Schwarzschild and
 Schwarzschild-(anti-)de Sitter black holes. For the Schwarzschild and
 Schwarzschild-anti-de Sitter backgrounds the stability of branes is
 shown analytically by means of a deformation technique. In contrast,
 for the Schwarzschild-de Sitter background  we demonstrate with the help of
 numerics that the brane is unstable (only) against the $s$-wave sector
 of perturbations. 
\end{abstract}

\vspace{.4cm}

\setcounter{footnote}{0}



\section{Introduction}

The system of an extended object such as a string and a brane
interplaying with a black hole appears in various contexts of
gravitational physics.
As a first stage of this direction, one may customarily 
neglect the thickness and the backreaction of brane into the
geometry, in which case the motion of brane is governed by Dirac-Nambu-Goto
action~\cite{Frolov1}. 
In spite of this test brane approximation, the system, which will be referred
to as a brane--black-hole system, is able to provide a rich variety of physically
intriguing results. Prime examples are the cosmological domain wall 
interacting with a primordial black hole, a domain wall pierced by a black hole~\cite{Chamblin:1999by}, and a probe D$q$-brane interacting
with background D$p$-branes, and so on. 
The final topic quoted above has attracted much interest 
from the viewpoint of the gauge/gravity duality, according to which 
D$p$/D$q$-branes describe a certain kind of phase transitions in gauge
theories~\cite{Mateos},  and its motion corresponds to the motion of quarks, mesons and baryons 
in the quark-gluon plasma~\cite{QGP}. 
It is also an interesting observation that the brane--black-hole system captures several aspects
of the `merger' of black holes appearing in the black-hole black-string
system in Kaluza-Klein spaces~\cite{Kol,Frolov3}.  
Besides these applications, the minimal surfaces, {\it i.e.},
the minimizer of the Dirac-Nambu-Goto action, is used as a technical
tool to prove some important theorems in general relativity such as the
positive mass theorem~\cite{Schon}.

In every context of the above applications of brane--black-hole system,
a natural question arises as to which equilibrium state is most stable
or what is the final state of dynamics such as 
gravitational scattering and capturing processes. 
The answer, of course, will depend on both the symmetries of
the background black-hole spacetime and the boundary conditions of the
brane at the asymptotic regions (infinity and horizons). 
As a specific example, let us consider a spherically symmetric
static black hole as a background geometry.   
A na\"ive expectation is that the final state of the dynamics of an
isolated brane--black-hole system is the configuration in which the
brane is trapped on the equatorial plane of the black hole,  because the
equatorial plane, which has the highest symmetries, will always be an
extremizer of the action and attractive force of the black hole seems to
put such a brane at the bottom of a concave potential.

As we will see soon, however, the problem turns out to be more subtle
practically than one expects as above. To assess the linear
stability of the brane at the equatorial plane, one is forced to 
evaluate a one-dimensional Schr\"odinger equation as occurred in a 
perturbative stability analysis of black hole. 
As known well, 
a careful treatment is required to conclude the stability 
even for higher dimensional static black holes possessing maximally symmetric 
horizons~\cite{KodamaIshibashi,IshibashiKodama}. In particular, the
presence of a cosmological constant and a charge/momentum makes the
potentials in the Schr\"odinger equation non-trivial and complexity 
increases as dimensionality becomes higher. The intricacy
of the effective potentials is the major obstacle for the analytic
study of stability.

In addition, we should
remind some facts known in the mathematical studies of minimal surfaces
in flat Euclidean space $\mathbf{E}^n$, which has a long history since
the pioneering works by Young and Laplace. The plane in $\mathbf{E}^3$,
which may be regarded as a counterpart of the brane on the equatorial
plane in our setting, is the only globally well defined single valued
graph of the minimal surface equation, which is known as the Bernstein
theorem, and its expected extension to general dimension is known
as the Bernstein conjecture. 
Although the spacetime dimensionality appears to play no r\^ole to prevent 
this conjecture,  this is  not the case~\cite{Bombieri}. 
The minimal surface need not be smooth necessarily
in arbitrary dimensions, and there exists a rather explicit example called
minimal cone for $n \geq 8$, which is curved and even singular at the
apex. 
Quite recently, it was pointed out that this failure of the Bernstein
conjecture and the existence of minimal cones had involved a variety of
gravitational and non-gravitational systems~\cite{GMM}, including the
stable (local) cone solution of brane in the brane--black-hole
system~\cite{Frolov3}.   

As far as the authors know, an initial attempt to address 
the stability of brane trapped on the
equatorial plane of a black hole was  made by 
Higaki~{\it et al.}~\cite{Higaki}, and this appears to be the only study in the
literature. They discussed 
the 4-dimensional Reissner-Nordstr\"{o}m-de Sitter black
and found a conclusive result that the electromagnetic charge and positive
cosmological constant tend to destabilize the brane. As mentioned 
above, however, the stability of minimal surfaces/branes in general
dimensions has non-trivial features, and the results of perturbation
analysis would shed light on this issue. Motivated by this and 
recent growing interest in multidimensional dynamics of gravity, 
this paper explores  the stability of branes trapped on their equatorial planes of
$d$-dimensional black holes ($d \geq 4$), including the case of
asymptotically anti-de Sitter black hole while switching off the
electromagnetic field for simplicity. 

The organization of this paper is as follows. In the next section, we
briefly review the covariant perturbation method of a test brane. 
In Sec.~\ref{sec:master} we specify the background spacetime and reduce
the covariant equation to a single eigenvalue problem in radial direction.
In Sec.~\ref{sec:stability} the (in-)stability of brane is examined. The
final section is devoted to discussions.  In Appendix~\ref{sec:bc}, 
a prescription to give the boundary condition of perturbations is given.
We use the geometrical unit  $c=G_d=1$, where $G_d$ is the
$d$-dimensional gravitational constant, and ($-,+,+,\ldots$) sign
convention. Our notation of the cosmological constant is that the
Einstein equation in $d$-dimension without a matter content takes the
form of $R_{\mu\nu} = 2\Lambda g_{\mu\nu} / (d-2) $, and we rewrite
$\Lambda = \epsilon (d-2)(d-1) / (2b^2)$ where $\epsilon=0$, $\pm 1$ and
$b$ $(>0)$ is a length scale.

\section{Covariant Perturbation Method}
\label{sec:pert}

Let us consider a timelike hypersurface embedded in the background
spacetime ($M,g_{\mu\nu}$) as $x^\mu = X^\mu(\xi^a)$, where $x^{\mu}$
($\mu,\nu = 0,1,\ldots,d-1$) and $\xi^a$ ($a,b=0,1,\ldots,d-2$) are the
coordinates of the spacetime and the hypersurface, respectively. 
The Dirac-Nambu-Goto action is
\be
	I = -\tau \int \dd^{d-1} \xi \sqrt{ - \gamma }\,,
\ee
where $\tau$ is the tension of the brane and $\gamma$ is the determinant of induced metric given by
\be
\gamma_{ab} := g_{\mu\nu} \pd_a X^\mu \pd_b X^\nu\,.
\ee
The equation of motion derived from the action is expressed as the vanishing trace of extrinsic curvature,
\be
	\gamma^{ab} K_{ab} = 0\,.
\label{eq:K=0}
\ee
Specifically, the brane motion is described  by minimal surfaces.

The perturbation of a background brane equilibrium can be thought of as
a test scalar field on the background brane regarded as a
$(d-1)$-dimensional submanifold ($\Sigma,\gamma_{ab}$). A systematic
covariant treatment was given in~\cite{Guven}. 
The quantitative measure of brane bending is expressed by projecting the
variation of the brane $\delta X^\mu $ 
into normal direction $n^\mu$ of the background equilibrium,
\be
	\Phi := n_\mu \delta X^\mu\,.
\ee
Varying the equation of motion (\ref{eq:K=0}), 
it turns out that the linear perturbation equations simplify to a scalar
field equation on the hypersurface,
\be
	\dalm^{(\gamma)} \Phi - \left( R^{(\gamma)} - h^{\mu\nu} R_{\mu\nu} \right) \Phi = 0\,,
\label{eq:eom-original}
\ee
where $h_{\mu\nu} := g_{\mu\nu} - n_\mu n_\nu $ is the projection tensor
onto the worldsheet.   
$ \dalm^{(\gamma)} = (-\gamma)^{-1/2} \pd_a  [ (-\gamma)^{1/2} \,
\gamma^{ab} \pd_b ]$
and $R^{(\gamma )}$ 
are the d'Alembertian operator  and the Ricci tensor 
associated with the induced metric $\gamma _{ab}$, respectively.

\section{Reduction to an Eigenvalue Problem}
\label{sec:master}

As a background spacetime, 
we consider the static, spherically symmetric $d$-dimensional spacetimes ($d\geq 4$) whose line element is
\be
	g_{\mu\nu} \dd x^\mu \dd x^\nu
	=
	- f(r) \dd t^2 + f(r)^{-1} \dd r^2
	+ r^2 \left( \dd \theta^2 + \sin^2 \theta \dd \Omega_{d-3}^2 \right)\,,
\label{eq:Sch}
\ee
where $\dd \Omega_{d-3}^2$ is the line element of a unit
($d-3$)-sphere. Let us consider the Dirac-Nambu-Goto ($d-1$)-dimensional
brane $(\Sigma,\gamma_{ab})$ with an $\mathrm{O}(d-2)$ invariance
embedded in the background~(\ref{eq:Sch}). Denoting the brane
configuration by $\theta=\theta(r)$, the induced metric reads
\be
	\gamma_{ab} \dd \xi^a \dd \xi^b
	=
	- f(r) \dd t^2 + \left[ f(r)^{-1} + r^2 \theta^{\prime 2}(r) \right] \dd r^2
	+ r^2 \sin^2 \theta \; \dd \Omega_{d-3}^2\,.
\ee
Without writing down the explicit form of Dirac-Nambu-Goto equation of
motion, one might deduce that the equatorial plane ($\theta =
\pi/2$) always extremalizes the action [since the equatorial plane is
geodesic in the $(r,\theta)$-plane]. This can be easily checked as
follows.  For the present brane configuration,  the Dirac-Nambu-Goto
action computes to give 
\be
	I
	=
	- \tau \Omega_{d-3} (t_2-t_1) \int \dd r\;
	\left( r \sin \theta \right)^{d-3} \sqrt{ 1 + f r^2 \theta^{\prime 2} }\,,
\ee
where $\Omega_{d-3}$ is the volume of a unit $(d-3)$-sphere, and
($t_2-t_1$) is a time interval. Then, varying this action, one obtains
an equation of motion,
\be
	\theta^{\prime\prime}
	+ \frac{r}{2} \left[ 2(d-2)f + rf^\prime \right] \theta^{\prime 3}
	- \frac{ d-3 }{ \tan \theta } \theta^{\prime 2}
	+ \left( \frac{ d-1 }{ r } + \frac{ f^\prime }{ f } \right) \theta^\prime
	- \frac{ d-3 }{ r^2 f \tan\theta } = 0\,.
\ee
Thus, $\theta(r) \equiv \pi/2$  obviously solves this equation, as we
desired to show. Hereafter, we shall concentrate on the brane located at
the equatorial plane and consider its linear perturbations.


The variables in equation~(\ref{eq:eom-original}) are separable by setting $
\Phi = r^{-(d-3)/2} \chi(r) e^{-i\omega t}Y(\Omega)$ and introducing the
tortoise coordinate $\dd r_\ast = \dd r/f$. Here, $Y(\Omega)$ is a
spherical harmonic  on a $(d-3)$-sphere (see, {\it
e.g.},~\cite{KodamaIshibashi}) satisfying 
\be
	\left[ \Delta^{(d-3)}  + \ell ( \ell + d-4 ) \right]Y = 0 \,,
\;\;\;
	\ell = 0,1,2,\ldots\,,
\ee
where $\Delta^{(d-3)}$ is the Laplace-Beltrami operator on the unit
$(d-3)$-sphere and $\ell $ is the multipole index.
It then follows that the wave equation reduces to the Schr\"odinger form,
\be
&&
	- \frac{ {\rm d}^2 \chi}{ {\rm d} r_\ast^2 }  + V(r) \chi
	=
	\omega^2 \chi\,,
\label{eq:eom}
\ee
with an effective potential,
\be
 V(r)=\frac{f}{r^2}\left[\ell (\ell+d-4)-(d-3)+\frac{(d-1)(d-3)}{4}f+\frac{d-1}{2}rf'\right]\,.
\label{V}
\ee

We have hitherto nowhere used Einstein's equations to derive the
perturbation equations. 
From now on, to make the discussion reasonably focused, 
we specialize to the cases where the background spacetime satisfies the
vacuum Einstein equation with or without a cosmological constant,
$R_{\mu\nu} = \epsilon(d-1)g_{\mu\nu}/b^2$, where $\epsilon=0$, $\pm 1$ and
$b$ $(>0)$ is a curvature radius. That is, we consider the case where
$f(r)$ takes the form,
\be
	f(r)
	=
	1 - \left( \frac{r_0}{r} \right)^{d-3} - \epsilon \left( \frac{r}{b} \right)^2 \,,
\label{eq:f}
\ee
where $r_0$ $(>0)$ is a constant corresponding to a mass of the black hole.
The case of $\epsilon=0$ and $\epsilon=+1(-1)$, respectively,
corresponds to the $d$-dimensional Schwarzschild black hole and
Schwarzschild-dS(AdS) black hole.\footnote{
More precisely, 
the metric function (\ref{eq:f}) for $\epsilon =1$ 
describes a Schwarzschild-dS black hole provided $r_0$ satisfies 
$r_0^{d-3}\le 2b^{d-3}(d-3)^{(d-3)/2}(d-1)^{-(d-1)/2}$. 
}
For the simplest Schwarzschild case, 
we first remark that the range of $r_*$ is complete 
$-\infty <r_*<\infty $ and hence the outside wedge of a black hole
is globally hyperbolic, indicating that the dynamics 
is uniquely determined by the initial data.  
The parallel will be inferred for the Schwarzschild-dS case 
if we recognize $r_*\to \infty $ 
at the cosmological horizon $r_c$ instead of infinity, 
where $r_c$ is the larger root of $f(r)=0$. 
In contrast to these two cases, the range of $r_*$ is incomplete,
 which we take $-\infty <r_*<0$,  for the Schwarzschild-AdS,  
implying that the outside region of a black hole ceases to be 
globally hyperbolic. This distinguished property demands a special care for 
the dynamics of $\chi$ in the $\epsilon=-1 $ case.

For the metric~(\ref{eq:f}),
the potential is written explicitly as 
\be
	V(r)
	=
	\frac{ f }{ 4r^2 }
	\left[
		4\ell(\ell+d-4) + (d-5)(d-3)
		+ (d-3)(d-1) \left( \frac{r_0}{r} \right)^{d-3}
		- \epsilon (d^2-1) \left( \frac{r}{b} \right)^2
	\right]\,.
\label{eq:pote}
\ee
Inspecting (\ref{eq:pote}), one finds 
a universal property that the potential $V$ is bounded below and vanishes 
at the event horizon $r_*\to -\infty$, independent of the sign of the
cosmological constant.   
However, behaviors of the potential $V$ away from the event horizon
is very sensitive to $\epsilon $.  
For $\epsilon \ge 0$, the potential vanishes at 
$r_*\to \infty $ (infinity for $\epsilon =0$ and 
cosmological horizon for $\epsilon =+1$). 
A distinct feature of a positive cosmological constant is that 
it will give a negative contribution to the potential [the last term in
equation (\ref{eq:pote})]. By virtue of this, $V$ approaches to zero from below 
as $r_*\to \infty $ and causes an instability, as we will 
show later. Whereas,  
the potential $V$ diverges at infinity in the case of $\epsilon =-1$. 
This illustrates that the negative cosmological constant acts as
a `confining box.'
In either case, the potential $V$ fails to be positive-definite
 in general.

In the rest of this paper, we examine the stability of the brane for each
background.

\section{Stability Analysis}
\label{sec:stability}

In order to conclude the linear stability, 
we must show that 
the Schr\"odinger equation $A\chi=\omega ^2\chi$ [equation~(\ref{eq:eom})]
admits no normalizable negative mode solutions $\omega^2<0$, where 
we have defined $A :=-\dd ^2/\dd r_*^2+V$. The operator 
$A$  is elliptic and identified as a Hamiltonian operator 
on a Hilbert space of square-integrable functions 
$L^2(r_*, {\rm d}r_*)$ on a static timeslice. 
To establish the stability,  
it is necessary to show that $A$ 
is a positive, self-adjoint operator with $L^2(r_*, {\rm d}r_*)$-norm.

Before embarking on the stability analysis, let us digress here and 
first focus on the dynamics of a scalar field $\chi $ in this background. 
Since the outside wedge is globally
hyperbolic for the Schwarzschild(-dS) spacetime, 
the dynamics of $\chi $ is well-posed in this region. 
Since the potential $V$ vanishes as
$r_*\to \pm \infty $, the only possible boundary condition
to obtain a normalizable solution is the Dirichlet boundary condition $\chi=0$
at $r_*\to \pm \infty $ (see Appendix~\ref{sec:bc}). Accordingly,  
the self-adjoint extension $A_E$ of $A$ is unique, that is to say, the
Hamiltonian $A$ is essentially self-adjoint. Hence all that remains to
solve for the stability is to show the positivity of $A_E$. 
This is a standard prescription to pursue the dynamics in 
globally hyperbolic spacetimes.

For the Schwarzschild-AdS case, on the other hand, 
the domain of outer communication is no longer 
globally hyperbolic.  This means that the ordinary Cauchy evolution 
determines a solution of evolution equations only within the domain of
influence for a given initial data slice.   
Nevertheless,  this difficulty can be cured for a {\it static} spacetime and 
it is always possible to find a sensible dynamics--and this is  essentially the unique recipe for
defining dynamics under quite reasonable conditions--beyond the domain
of dependence of initial data slice~\cite{IshibashiWald}.   
To define a unitary dynamics throughout the non-globally hyperbolic
static spacetime, we need a self-adjoint extension of $A$,  
which may or may not be unique. As is well known,
choosing a self-adjoint extension is equivalent to choosing a 
boundary condition.

Let us determine possible boundary conditions for $\epsilon=-1$. 
Since the potential vanishes at the horizon, 
the normalizability singles out a unique boundary condition therein.
Henceforth, our primary concern is the boundary condition at infinity.  
Near infinity $r_*\sim -b^2/r\sim 0$, $A$ behaves as
\begin{align}
 A\sim -\frac{\dd ^2}{\dd r_*^2}+\frac{d^2-1}{4r_*^2},
\end{align}
this yields the asymptotic solution,
\begin{align}
 \chi \sim C_1(r_*^{(d+1)/2}+\cdots )+C_2 (r_*^{-(d-1)/2}+\cdots ).
\end{align}
The normalizability requires $C_2=0$, {\it viz}, $V$ is in the 
limit point case at infinity. Therefore, 
the self-adjoint extension of $A$ is found to be unique even for $\epsilon=-1$. 
This observation is central to the stability discussion developed below.

Given a self-adjoint extension $A_E$ of $A$, let us next 
move on to the issue on the positivity of $A_E$. 
If the potential $V$ happens to be positive, $A_E$
corresponds to a Friedrichs extension and the stability 
against linear perturbation immediately follows~\cite{Wald}.
As we have seen, however, the potential function $V$, equation
(\ref{V}), is not positive-definite, so it is far from obvious 
whether $A_E$ has positive eigenvalues. 
To see the positivity of $A_E$,  it is of great benefit to 
employ the idea proposed by
Ishibashi and Kodama~\cite{IshibashiKodama}.  
Following their algorithm,  the potential in the
Schr\"odinger equation can be `deformed' in such a way that a
newly deformed potential is positive-definite.  
Let $S$ denote an arbitrary function of $r$ and 
\begin{align}
\hat D = \frac{\rm d}{{\rm d}r_*} -S
\end{align}
be a derivative operator. 
Straightforward calculation shows that 
\begin{align}
 (\chi, A\chi)_{L^2} = -\left[ \bar{\chi} \hat{D} \chi\right]_{\rm boundary }+\int
{\rm d}r_* \left(\left| \hat D\chi\right|^2 +V_S
 \left|\chi\right|^2\right)\,,
\label{deform}
\end{align}
 and
\begin{align}
 V_S= V-f\frac{\rm d}{{\rm d}r}S-S^2\,.
\end{align}
It deserves to remark that this expression is formal because 
there is in general no guarantee that the boundary term and the integrand
are both finite. If one can show that the 
boundary term  vanishes and if one can find an appropriate $S$
that makes $V_S$ positive, 
the Hamiltonian operator $A$ turns out to be a positive operator, which 
has at least one positive self-adjoint extension corresponding to  the 
Friedrichs extension.
Utilizing this formula, we are thereby able to demonstrate the positivity of 
$A_E$, {\it i.e.}, the stability of test brane perturbations.

\subsection{Schwarzschild(-AdS) Background}
\label{sec:Sch}

Let us begin by the analysis of $\epsilon\le 0$. 
As we have shown that the boundary condition is necessarily
of Dirichlet-type, it then follows that the boundary term in 
equation~(\ref{deform}) indeed  drops off and
$V_S \ge 0$ implies a stable dynamics. 
We can accomplish this by the choice 
\begin{align}
 S = \frac{d-3}{2r}f\,,
\end{align}
giving rise to a positive potential for any $\ell \ge 0$, 
\begin{align}
 V_S= \frac{f}{r^2}\left[\ell (\ell +d-4)-\epsilon (d-1)\left( \frac{r}{b} \right)^2\right]\,.
\end{align}
Then, the test brane perturbation in Schwarzschild(-AdS) black hole 
is stable for any $d\ge 4$ dimensions.\footnote{One can show that the string motion
in the background of BTZ black hole ($d=3, \epsilon =-1$)~\cite{BTZ} is also stable. 
}

\subsection{Schwarzschild-dS Background}
\label{sec:Sch-dS}

According to~\cite{Higaki}, the four-dimensional 
Schwarzschild-dS black hole exhibits an instability only for the 
$\ell =0$ mode. We now show that this result continues to be true for all $d\ge 4$. 

As the domain surrounded by event and cosmological horizons is globally
hyperbolic, equation~(\ref{deform}) receives no boundary contribution.  
We find that 
\begin{align}
  S= \frac{ d-1 }{2r}f
\end{align}
leads to the deformed potential
\begin{align}
  V_S = \frac{ (\ell -1)(\ell+d-3)f }{ r^2 }\,.
\label{defV2}
\end{align}
So the system  is stable in $d\ge 4 $ dimensions 
against for all $\ell \ge 1$ modes. 
A possible unstable mode exists only for the 
$s$-wave.\footnote{It is worthwhile to emphasize that
equation  (\ref{defV2}) can be derived from equation  (\ref{V})
without any reference to the particular form of $f$. 
Thus, all the spherically symmetric static black-hole type backgrounds of the
form~(\ref{eq:Sch}) admit a stable brane dynamics for $\ell \ge 1$ modes.}

We will next show numerically that the $s$-wave actually excites an
 unstable mode. To perform numerics, we have to 
 make dimensionless quantities.  Here, 
the normalization by $r_0$ seems not so useful 
because it is a measure of the mass, which has the dimensional dependence
as $r_0^{d-3}\sim G_dM$. 
To compare black holes in different spacetime dimensions, 
it is more advantageous, instead of $r_0$ and $b$, to use $r_c$
and $r_h$, which are, respectively, the radii of the cosmological
horizon and the event horizon.
Simple calculations show that they are related by
\begin{align}
r_0^{d-3}=\frac{r_h^{d-3}r_c^{d-3}(r_c^2-r_h^2)}{r_c^{d-1}-r_h^{d-1}}\,,\qquad
b^{-2} =\frac{r_c^{d-3}-r_h^{d-3}}{r_c^{d-1}-r_h^{d-1}}\,.
\label{para}
\end{align}
One may recognize immediately that 
the background  becomes the Schwarzschild black hole
as the ratio $r_c/r_h$ tends to infinity.

We show the values of growth rate $\sigma:= {\rm Im}(\omega)  $ of 
perturbations in typical dimensions in Table~\ref{table:sigma}.\footnote{The proof of $\omega $ being pure imaginary is relegated to Appendix. The numerical results presented in Table~\ref{table:sigma} were obtained by two independent shooting calculations. However, the numerical credibility becomes worse as the dimensionality increases, $d \gtrsim 10$, in particular for large $r_c/r_h$.
It would be interesting to apply the matched asymptotic expansion method for the large-$d$ limit~\cite{Asnin:2007rw}, by which one may be able to determine the growth rate analytically.}
When the ratio $r_c/r_h$ is fixed, the growth rate $\sigma r_h$ becomes 
higher as the dimension gets higher.
The brane on the equatorial plane of the higher 
dimensional Schwarzschild-de Sitter black hole is more unstable 
than that of the lower dimensional one  
 with  the identical value of ratio $r_c/r_h$.
It can be understood by the form of potential in each dimension.
When the spacetime dimension is fixed, the growth rate $\sigma r_h$
becomes lower as the ratio $r_c/r_h$ gets bigger.
In other words, smaller cosmological constant reduces the instability of the system.
This tendency is compatible with the fact that the system in the case of
the Schwarzschild black hole background is stable.

To sum up, we found that the positive cosmological constant 
strengthens the instability of branes on the equatorial plane 
for the arbitrary dimensional Schwarzschild-de Sitter black holes.
We can also say that the higher dimensional background 
has a destabilization effect of the system.

\begin{table}[t]
	\begin{center}
\caption{Numerical values of the growth rate $\sigma = \mathrm{Im}( \omega )$ of the
	 $\ell=0$ perturbation of brane in the Schwarzschild-de Sitter
	 background. The horizon radius $r_h$ is set to unity.}
\label{table:sigma}
\vspace{6pt}
\setlength{\tabcolsep}{17.3pt}
\footnotesize{
\begin{tabular}{cccccc}
\hline \hline
 $r_c$ & $d=4$ & $d=5$ & $d=6$ & $d=10$ & $d=11$
\\
\hline
 $2$ &	$1.332 \times 10^{-1}$ & $2.429 \times 10^{-1}$ & $3.277 \times 10^{-1}$ & $4.762 \times 10^{-1}$ & $4.865 \times 10^{-1}$
\\
  $20$ & $4.435 \times 10^{-2}$ & $4.957 \times 10^{-2}$ & $4.997 \times 10^{-2}$ & $5.00 \times 10^{-2}$ & $5.00 \times 10^{-2}$

\\ 
  $100$ & $9.750 \times 10^{-3}$ & $9.996 \times 10^{-3}$ & $1.000 \times 10^{-2}$ & $1.00 \times 10^{-2}$ & $1.00 \times 10^{-2}$ 

\\  
\hline\hline
	\end{tabular}
}
	\end{center}
\end{table}

\section{Discussions}
\label{sec:disc}

We have seen that the branes rest at the equatorial plane of the
$d$-dimensional ($d \geq 4$) Schwarzschild and Schwarzschild-AdS black
holes are stable for the linear perturbations, while for the
Schwarzschild-dS background the brane is unstable against the $s$-wave
perturbation. The covariant perturbation equation of brane is translated
into the the eigenvalue problem in the Schr\"odinger form after the
separation of variables. Then, the stabilities were proved by showing
the positivity of the Hamiltonian operator by means of a deformation
technique. In the asymptotically de Sitter case, the positivity of
Hamiltonian operator only for the non-spherical perturbations 
($\ell \geq 1$) was shown, which means that one has to solve the eigenvalue
problem explicitly to pursue the behavior of spherical mode. 
To this end, we numerically solved the equation for
the spherical perturbation and showed that the mode is an unstable one in
every dimension, growing exponentially in time. 
One can recognize this coming from the fact that large  $\ell $ 
has the stabilizing effect. 
The dangerous modes originate from the lower angular eigenvalues since 
the effective potential becomes higher as $\ell $ increases. 
This result is
reminiscent of the fact that the Gregory-Laflamme instability of black
branes exists only in the $s$-wave sector~\cite{GL}. 

For definiteness of our argument, we have limited our consideration to 
a neutral black hole as a background black hole. 
The inclusion of electromagnetic charge is straightforward.  
The electromagnetic charge will
leave a negative imprint on the effective potential~(\ref{V})
as in an analogous fashion in four dimensions~\cite{Higaki}, where
an instability was explicitly shown. 
Thus there seems no compelling reason to stabilize the brane motion
in the charged case. 
As we said, this instability occurs only for the $s$-wave in
any spacetimes of the form (\ref{eq:Sch}) such as a black hole
in different gravitational theories.

In this paper, we have ignored the backreaction of brane self-gravity
and the thickness of brane, both of which cannot be negligible in
general situations. If one takes into account the backreaction, it is
known that the domain wall exhibits repulsive gravitational
fields~\cite{Vilenkin}. Regarding the thick branes interplaying with black
holes, several equilibrium solutions were constructed numerically or
analytically~\cite{Morisawa, Rogatko:2003ve}. The introduction of the thickness is known
to result in the expulsion of branes/domain walls by an
extremal/near-extremal black hole. Thus, it would be interesting to see
how the competition between the attractive and repulsive forces
appearing in more realistic modelings of brane--black-hole system
affects the stability of the brane examined in this paper. 

While we have restricted ourselves to the linear stability of brane and
found, for example, the stability in the Schwarzschild background, it is
known that a certain sort of non-linear perturbations to the
Dirac-Nambu-Goto brane in the Schwarzschild background, which mimic the
recoil of black hole due to the Hawking emission, evolves and results in
a reconnection of the bent brane, corresponding to the eventual escape
of black hole into the bulk~\cite{Flachi}.  
It is noted that our results are not inconsistent with those
in~\cite{Flachi} since the asymptotic boundary conditions are different.
In a similar vein, the existence of a critical escape velocity of the 
black hole from the brane was 
investigated in~\cite{Flachi:2006ev}. Interestingly, the existence of
critical escape velocity depends on the number of codimensions. Although
we have been concerned only with the codimension-one branes in this
paper, it would be interesting to investigate how the number of
codimensions affects the stability of the brane, which can be
examined with the covariant perturbation methods developed
in~\cite{Larsen}.

\subsection*{Acknowledgments}
U.M. would like to thank Barak Kol for conversations.
We are grateful to Antonino Flachi and Hideki Maeda for useful disucussions and comments.
Works of U.M.~are supported by the Lady Davis Fellowship,
by the Israel Science Foundation Grant (No.~607/05), and by the DIP Grant 
(No.~H.52).

\appendix

\section{Boundary Conditions}
\label{sec:bc}

Here, we assemble several basic facts about the behaviors of physical
perturbations in the case of $\epsilon \geq 0$, which are used in this
paper and might be helpful for readers. 

For the Schwarzschild case ($\epsilon=0$), $r_\ast \to -\infty$ as $r
\to r_h$ ($r=r_h$ is the event horizon) and $r_\ast \to +\infty$ as $r
\to +\infty$. For the Schwarzschild-dS case ($\epsilon=+1$), $r_\ast \to
-\infty$ as $r \to r_h$ again and $r_\ast \to +\infty$ as $r \to r_c$
($r=r_c$ is the cosmological horizon). 
One can see that the potential (\ref{eq:pote}) vanishes as $r_\ast \to \pm \infty$.
Thus, the linearly independent asymptotic solutions are the plane waves, $e^{\pm i\omega r_\ast}$.
Since we assume the time dependence of $\chi \propto e^{-i\omega t}$,
two modes $e^{-i\omega r_\ast}$ and $e^{i\omega r_\ast}$ correspond to
the ingoing and outgoing waves, respectively. 
Hence, we impose the following boundary conditions on a physical mode, $
\chi \to e^{ \pm i\omega r_\ast } $ as $r_\ast \to \pm \infty$. 

As far as an unstable mode, corresponding to $ \mathrm{Im}(\omega) > 0
$, is concerned, we can show that  $\omega$ is pure imaginary as
follows. Multiplying the complex conjugate of $\chi$, $\bar{\chi}$, with
equation~(\ref{eq:eom}), and integrating it, we have 
\be
(\chi, A\chi)_{L^2}=
	\int_{-\infty}^{+\infty}
	\left(
		\left| \frac{ \dd\chi }{ \dd r_\ast } \right|^2
		+
		V | \chi |^2 
	\right) \dd r_\ast
	-
	\left[
		\bar{\chi} \frac{\dd \chi}{\dd r_\ast}
	\right]_{-\infty}^{+\infty}
	=
	\omega^2 \int_{-\infty}^{+\infty} | \chi |^2 \dd r_\ast\,.
\label{eq:int1}
\ee
Since the surface term in equation~(\ref{eq:int1}) vanishes due to the
exponential suppression at infinity, the all integrals in
equation~(\ref{eq:int1}) are real. Thus, $\omega^2$ must be real, and
furthermore $\omega$ must be pure imaginary [since we are looking for
the unstable mode, $\mathrm{Im}(\omega)>0$]. 
Therefore, we can write as $\omega = i \sigma$ $(\sigma > 0)$ and
equation~(\ref{eq:int1}) reads 
\be
	- \sigma^2
	=
	\int_{-\infty}^{+\infty}
	\left(
		\left| \frac{ \dd\chi }{ \dd r_\ast } \right|^2
		+
		V | \chi |^2 
	\right) \dd r_\ast
	\Bigg/
	\int_{-\infty}^{+\infty} | \chi |^2 \dd r_\ast\,.
\label{eq:int2}
\ee
What we can see from this equation is that if the potential $V(r)$ is
positive definite, the right hand side is positive and
equation~(\ref{eq:int2}) makes no sense, implying the absence of unstable
mode. 
In terms of $\sigma$, the asymptotic boundary conditions mentioned above
are rewritten as $ \chi \to e^{\pm \sigma r_\ast} $ as $r_\ast \to \mp
\infty$, which are used in the numerical analysis.



\end{document}